\documentclass[12pt,letterpaper]{article}
\usepackage[top=1in,bottom=1in,left=1in,right=1in]{geometry}
\usepackage[utf8]{inputenc}
\usepackage{hyperref}
\usepackage{amsmath}
\usepackage{amssymb}
\usepackage[square,comma,numbers]{natbib}
\usepackage{titlesec}

\titlespacing*{\section}
{0pt}{1.5ex plus 1ex minus .2ex}{0.4ex plus .2ex}

\title{Gravity and Light: Combining Gravitational Wave and Electromagnetic Observations in the 2020s}

\date{\today}

\begin{document}

%\maketitle

\LARGE
\begin{center}
    Gravity and Light: Combining Gravitational Wave and Electromagnetic Observations in the 2020s
\end{center}
\normalsize
%\bigskip
%\bigskip
%\bigskip

\footnotesize
\noindent
{\bf Thematic areas:
3. Stars and Stellar Evolution
4. Formation and Evolution of Compact Objects
7. Cosmology and Fundamental Physics
8. Multi-Messenger Astronomy and Astrophysics}\\

\noindent
{\bf Principal Author:}\\
Ryan~J.~Foley\\
UC Santa Cruz\\
foley@ucsc.edu

\bigskip

\noindent
{\bf Co-authors:}\\
K.\ D.\ Alexander$^{1}$,
I.\ Andreoni$^{2}$,
I.\ Arcavi$^{3}$,
K.\ Auchettl$^{4}$,
J.\ Barnes$^{5}$, 
G.\ Baym$^{6}$,
E.\ C.\ Bellm$^{7}$,
A.\ M.\ Beloborodov$^{5}$, 
N.\ Blagorodnova$^{8}$, 
J.\ P.\ Blakeslee$^{9}$
% ...

\noindent
$^{1}$Northwestern,
$^{2}$Caltech,
$^{3}$Tel Aviv,
$^{4}$DARK,
$^{5}$Columbia,
$^{6}$UIUC,
$^{7}$UW,
$^{8}$Raboud,
$^{9}$Gemini,\\
% ...
\normalsize
\bigskip

\noindent
{\bf Abstract:}\\
As of today, we have directly detected exactly one source in both gravitational waves (GWs) and electromagnetic (EM) radiation, the binary neutron star merger GW170817, its associated gamma-ray burst GRB170817A, and the subsequent kilonova SSS17a/AT~2017gfo.  Within ten years, we will detect hundreds of events, including new classes of events such as neutron-star--black-hole mergers, core-collapse supernovae, and almost certainly something completely unexpected.  As we build this sample, we will explore exotic astrophysical topics ranging from nucleosynthesis, stellar evolution, general relativity, high-energy astrophysics, nuclear matter, to cosmology.  The discovery potential is extraordinary, and investments in this area will yield major scientific breakthroughs.  Here we outline some of the most exciting scientific questions that can be answered by combining GW and EM observations.

\pagenumbering{gobble}
\newpage
\pagenumbering{arabic}

\section{The Dawn of Multi-Messenger Astronomy}

On August 17, 2017, the LIGO-Virgo collaboration (LVC) detected, for the first time, a gravitational wave (GW) signal (GW170817) from two inspiralling neutron stars \citep[NSs;][]{Abbott17:gw170817}.  At the same time an electromagnetic (EM) counterparts, a short gamma-ray burst (SGRB; GRB170817A) and a kilonova (called SSS17a or AT~2017gfo), was detected at all wavelengths \citep{Abbott17:grb, Coulter17, Hallinan17, Troja17}.  With a single event, we were able to address fundamental questions in general relativity \citep{Pardo18}, nuclear matter \citep{Abbott17:gw170817}, cosmology \citep{Abbott17:h0, Creminelli17, Ezquiaga17}, nucleosynthesis \citep{Kasen17}, and astrophysics \citep[e.g.,][]{Cowperthwaite17:gw, Drout17, Kasen17, Kasliwal17, Kilpatrick17:gw, Pian17, Smartt17}.

In April 2019, LVC will start its third observing run, O3, at increased sensitivity.  O3 will last about a year (roughly through March 2020).  LVC expects to detect between 1 and 50 binary NS (BNS) mergers during O3 with a median estimate of about 10 events \citep{Abbott16:review}.  There is also the possibility of discovering the first neutron-star--black-hole (NSBH) merger, which should produce an EM counterpart similar to (but likely fainter than) those of BNS mergers.  Toward the end of O3, KAGRA may join the network, and with it, we will have our first 4-detector events.  In 2021, LVC (with KAGRA) is expected to start another observing run at design sensitivity \citep{Abbott18:future}.  LIGO India should join the network in 2025.  Over the next decade, we will move from having a single event seen in both gravitational waves and light to a sample of hundreds.

As this new field quickly evolves, we are poised to use gravitational waves and light in concert to answer long-standing questions in nucleosynthesis, stellar evolution, nuclear matter, and cosmology.  {\bf We are confident that the combination of gravity and light will allow us to make major strides in several vital areas, including generating surprises.}

\section{What are the Origins of the Heaviest Elements?}

The light curves and spectra of SSS17a indicate that up to 0.05~M$_{\odot}$ of $r$-process material was created in and ejected from the merger.  Combined with the BNS merger rate, BNS mergers could produce all of the $r$-process material in our Galaxy {\it if} SSS17a is typical \citep{Cowperthwaite17:gw, Kilpatrick17:gw, Rosswog99, Rosswog18, Smartt17, Tanvir17}.  However, the total yield per kilonova depends on the mass ratio of the binary and the total remnant mass.  Studies of emission from SGRBs indicate that GW170817 produced more $r$-process material than the typical SGRB \citep{Ascenzi18, Gompertz18}.

Even in the case of SSS17a, the exact amount of $r$-process material, and particularly ``heavy'' or ``3rd-peak'' material, is still quite uncertain \citep{Wu18}.  The optical and infrared radiation in kilonovae can arise from a variety of ejecta components and even radioactive species in the merger, tidal tails, and accretion disk winds \citep{Kasen17, Metzger18}, and precise mass estimates require a full understanding of the emission mechanisms and careful accounting. For SSS17a it is still debated whether the color evolution was caused by (1) decay of $r$-process nuclei in ejecta with different compositions \citep{Kawaguchi18, Villar17}; (2) decay of material from a single origin \citep{Smartt17, Waxman18}; or (3) energy sources such as shock cooling \citep{Piro18}, remnant winds \citep{Yu18}, or central engine activity \citep{Kisaka16}. Ultraviolet through infrared light curves of a statistical sample of kilonovae will allow us to distinguish between models and address key theoretical questions.

An intriguing path toward measuring the amount of heavy $r$-process material is through careful examination of the kilonova light curve at late times.  At early times, many radioactive species contribute to the luminosity and the combination of all decays results in a power-law-dependent heating rate.  At later times, as the majority of species decay away, the kilonova light curve is powered by at most a handful of long-lived radioisotopes.  Theorists have identified $^{254}$Cf, a heavy $r$-process element with a half-life of 60.5~days \citep{Wu18, Zhu18}, as being a possible dominant energy source at late times.  Measuring the kilonova luminosity two months after the merger will reveal the amount of $^{254}$Cf generated in the merger \citep{Kasliwal19}.  Combined with the robust solar abundance pattern and the early-time data, we can then accurately determine the relative abundance of light-to-heavy $r$-process material created in BNS mergers.

Perhaps the most compelling path toward further understanding element creation is through spectral modeling \citep{Kasen17}.  Currently, atomic line lists are insufficient for directly determining the abundance of individual elements in kilonova spectra \citep{Kasen13}.  Improvements in this area will require significant laboratory work, but it would be extremely beneficial to this science area.  Better spectral modeling will also strengthen our constraints on the dynamics of the ejecta, especially once we obtain a nebular spectrum with JWST or an ELT.

\section{What are the Details of the Merger and Explosion?}

Numerical relativity can precisely reproduce the waveform for the inspiral before a BNS merger \citep{Abbott17:gw170817}.  However, LVC was unable to detect the actual merger for GW170817 since its frequency increased to a point where its detectors lose sensitivity.  As a result, we must rely on numerical simulations to understand the actual merger process \citep{Foucart19, Kiuchi09, Paschalidis15, Rosswog02:merger, Ruiz18}.  These simulations generically show that the NSs should be (partially) tidally disrupted as they merge.  Depending on the individual NS masses and the mass ratio of the system, there is a broad range of outcomes.  The ejecta masses, ejecta velocities, number of tidal tails, and duration/size of an accretion disk all depend on the details of the progenitor system.

The merger product could promptly collapse to a BH \citep{Bauswein13} or could be an accreting hypermassive NS (HMNS) that exists for several milliseconds before collapsing to a black hole \citep{Shibata17}.  If the latter, a significant neutrino flux will exist during this time, which can irradiate the disk wind, changing the electron fraction and thus the final composition \citep{Drout17, Evans17, Just15, Kilpatrick17:gw, Metzger14}.  The size of the disk and the lifetime of the HMNS both depend on the equation-of-state (EoS) for nuclear matter \citep{Hotokezaka13}.  Since the composition of the ejecta affects the EM observables, one can hope to invert the problem to understand the immediate aftermath of the merger \citep{Bauswein17}.

However, additional physics may complicate the picture.  In particular, we expect a relativistic jet to be launched from at least some BNS mergers.  If the jet escapes, it should heat a cocoon of material surrounding the jet that can produce additional early-time emission.  Shock cooling, a long-lived engine, and additional disk-wind emission could all contribute to the early-time flux.  Detailed modeling of these processes and comparison to the nonthermal emission at $\gamma$-ray, X-ray, UV, and radio wavelengths, especially observations within the first day after merger as well as months later, should differentiate between scenarios \citep{Alexander18, Arcavi18, Ghirlanda18, Lazzati18, Mooley18:late}.

The combination of GW and EM data will be particularly useful for compact binary coalescences.  The GW signal precisely determines the merger time, removing a key variable from modeling the EM data.  For these events the GW data precisely measures the binary chirp mass, but the mass ratio is not as well constrained.  GW data also set upper limits to the tidal deformations and thus the radii of NSs.  EM data complement these analyses, as these data in combination with numerical relativity simulations place lower limits on the tidal deformation and more stringent constraints on the mass ratio. Independently, the EM signal depends on the maximum mass of NSs \citep{Margalit17}, a unique look at the EOS at the center of the NS.  In this way, combined GW and EM data allow for a better understanding of the outcomes and physical processes of compact binary coalescence mergers \citep[e.g.,][]{Coughlin18:mma, De18, Radice18}

\section{What Systems Create Compact-Object Mergers?}

GW observations of binary BHs have revealed the population of massive stellar-mass BHs \citep[20--50~M$_{\odot}$;][]{Abbott19:sample}, raising questions about which stars produce the BHs, how such systems form, and what the minimum and maximum masses are for stellar-mass black holes.  Similar questions exist for systems that include at least one NS, but with even fewer data.  While we have observed a limited number of BNS systems in the Milky Way and have some constraints from short gamma-ray bursts (SGRBs), both populations must be biased subsamples of the full population of compact-object binaries.

As we build up statistical samples of compact-object mergers, we will be able to examine their demographics, revealing details of the underlying population.  As an example, the SGRB host-galaxy population is skewed toward star-forming galaxies, but a significant fraction still resides in passive galaxies \citep[e.g.,][]{Fong13:sample, Prochaska06}.  This implies a delay-time distribution (the time from star formation until merger) is peaked toward short delay times, but has a long tail (if gravitational radiation is the slowest step, one expects a rate proportional to $t^{-1}$, consistent with the observations).  Similarly, a sizable fraction of SGRBs are found far from their host galaxies or have no associated host galaxy \citep{Fong13:loc}, indicating that these systems were expelled from their birth sites by supernova (SN) kicks.  But it is unclear how much the population of SGRBs, especially those with localizations sufficient to determine a host galaxy, overlap with BNS mergers.

We will be able to independently determine the delay-time distribution and offset distribution for a sample originating from GW detections.  With detailed analyses of the host environments, we will generate a more explicit link to BNS progenitor systems \citep{Blanchard17, Hjorth17, Levan17, Palmese17, Pan17:gw}.

We will also be able to directly compare the physical properties determined from the GW signal with both kilonova properties and population statistics.  For instance, one might expect systems with large mass ratios to be disrupted if there is a large SN kick and therefore all such systems are located within two half-light radii of their host galaxy.

While answering questions about the underlying population, we will also determine if every BNS merger produces a GRB, if SGRB samples are biased against significantly kicked mergers, and if BNS and NSBH mergers produce similar gamma-ray observables.

\section{How Fast is the Universe Expanding?}

New, independent measurements of $H_{0}$ are especially important since there is currently a 3.8-$\sigma$ disagreement between the local value \citep{Riess18:gaia} and that of the inverse-distance ladder \citep{Planck18}.  As the $H_{0}$ tension is one of the most significant open questions in cosmology and could reveal new physics, including non-$\Lambda$ dark energy, it is essential to use as many reliable techniques as possible.

Unlike most astrophysical sources, the waveform of GWs directly encodes the luminosity distance \citep{Schutz86}.  However, once an EM counterpart is found, we can independently measure a redshift from a host galaxy or the transient itself (which could be especially crucial for events kicked far from their host galaxy).  With both the luminosity distance and redshift, we can measure $H_{0}$.

With GW170817, we made the first standard siren measurement of $H_{0}$ \citep{Abbott17:h0}, finding a value consistent with both the local and inverse distance ladder values.  The initial analysis only used the position of the source and the redshift of its host galaxy.  The constraint on $H_{0}$ was limited by the degeneracy between $H_{0}$ and the inclination angle of the binary system.  The inclination angle measurement was further improved by radio observations of the jet \citep{Mooley18:super}, significantly reducing the uncertainty on $H_{0}$ \citep{Hotokezaka18}.

% (Fig.\ \ref{f:hubble}).

%\begin{wrapfigure}{r}{0.5\textwidth}
% \vspace{-20pt}
%  \begin{center}
%    \includegraphics[angle=0, width=0.48\textwidth]{H0-cosiota}
%  \end{center}
% \vspace{-23pt}
% \caption{Posterior probability of $H_{0}$ and inclination angle ($\iota$) from the joint GW-EM analysis (blue contours).  The 1- and 2-$\sigma$ contours are in black.  Best-fit values of $H_{0}$ and 1- and 2-$\sigma$ error bands are also displayed from Planck \citep{Planck16} and SH$_{0}$ES \citep{Riess16}.  EM constraints on the inclination angle can improve the measurement of $H_{0}$ \citep[e.g.,][]{Hotokezaka18}.  Figure from \citep{Abbott17:h0}.}\label{f:hubble}
%\vspace{-12pt}
%\end{wrapfigure}

As new events are added to the sample, the precision scales roughly as $\sqrt{N}$ \citep{Chen18:siren, Chernoff93, Dalal06, Nissanke13}.  The exact improvement depends on the details of the system, where some objects with smaller distance uncertainties or very well-constrained inclination angles will be much more valuable than others.  Simple predictions indicate that we can make a 3\% measurement of $H_{0}$ with 40 GW events with counterparts \citep{Chen18:siren}, which should be achievable within the next few years.

%EM data can significantly reduce the uncertainty in the inclination angle, directly improving the measurement of $H_{0}$.  For GW170817, the GW-only $H_{0}$ (along with the EM position and host-galaxy redshift) precision was 16\% \citep{Abbott17:h0}; including the constraint on the jet viewing angle based on the time of peak brightness for the non-thermal emission, the precision improved to 13\% \citep{Guidorzi17}; including the superluminal motion of the jet with VLBI measurements, the precision improved to 6.8\% \citep{Hotokezaka18}.  Those observations were extraordinary --- we do not expect to observe non-thermal emission for all events, let alone see superluminal motion of the jet --- and future events will likely have relatively limited data.  Luckily, the early-time color, dictated by the relative flux from the blue lanthanide-poor and red lanthanide-rich ejecta, is highly viewing-angle dependent \citep{Kasen17}.  With even a few data points, we may be able to improve inclination constraints.

%To further improve the constraints using only the kilonova emission, new theoretical models are necessary.  We will generate new merger models and produce new time-evolving spectral-energy distributions with radiative transfer codes (see above).  Comparing these models to the data, we will attempt to further constrain the inclination angle of each event.

Finally, once the population has increased, we can use properties of the sample to further improve all measurements.  Specifically, BNS mergers should be randomly oriented \citep[although we need to pay careful attention to selection effects;][]{Mandel18}.  With a hierarchical model, we can enforce that our sample has an appropriate inclination distribution.  This effectively reduces the possible inclination angles for otherwise poorly constrained events.

While systematic uncertainties are currently subdominant, they may become important at the few percent level.  Further improving the calibration of the GW strain signal and studies of the merger and kilonova emission will be critical to harness the full statistical sample that we will have in 10 years.

With improved GW facilities, we will be able to detect events sufficiently far  to measure $H(z)$ {\it directly}.  Future facilities such as LIGO Voyager and the proposed Cosmic Explorer should be able to detect BNS mergers to redshift $z \approx 0.4$ and $z \approx 7$, respectively\footnote{ \url{https://dcc.ligo.org/public/0150/T1800084/005/gw\_horizons.pdf}}.  Combined with sensitive space-based NIR observatories such as {\it JWST}, {\it WFIRST}, \textit{LUVOIR}, and ground-based ELTs, we could directly track the expansion history of the Universe to the age of deceleration, providing an independent probe of dark energy \citep{Dalal06}.

\section{What Will We Learn from Other Sources of Gravitational Radiation?}

During the process of core collapse, stellar cores are expected to generate relatively strong gravitational radiation \citep{Kotake13}.  While not nearly as loud as BNS mergers, the LVC is already sensitive to Milky Way supernovae \citep[with some extreme models being much louder and possibly detectable in other galaxies;][]{Abbott16:sn}.  Such an SN in the Milky Way would be especially interesting as we would also expect to detect thousands of associated neutrinos \citep{Scholberg12}.

Although the GW signal from core collapse should be more complicated than binary coalescence, the waveform would encode critical information about the core-collapse process, the proto-NS, and the SN mechanism \citep{Logue12, Powell18}.

As Galactic SNe are relatively rare (there is perhaps a 10\% chance of one occurring in the 2020s), it is possible that GW detectors will improve quickly enough that the first GW detection from an SN will be in another galaxy.  Future upgrades and new facilities could reliably detect SNe in the Virgo cluster.  As there are a handful of core-collapse SNe within 20~Mpc each year, we could have a sample of several by the end of the decade.  With such a sample, we would be able to determine if there are multiple explosion mechanisms and if the kinematics of the core is related to the success of an SN, the mass of the star, or the details of the resulting explosion.

While we {\it expect} to eventually detect SNe with GWs, there could be GW sources not yet theorized.  Every other time astronomers have opened a new window to the Universe, there have been new physical phenomena observed.  For GWs, the discovery space is particularly large, increasing the anticipation and expectation for something new.  EM observations will be critical to localize and characterize any such source.

\section{Recommendations}

The combination of gravitational wave and electromagnetic observations will be one of the most fruitful scientific endeavors of the next decade.  While this scientific field is still in its infancy, the scientific return for a wide range of fields, extending beyond astrophysics, is enormous.  To take full advantage of this opportunity, we make the following recommendations.
\begin{itemize}
  \setlength\itemsep{-3pt}
    \item Increase the number, sensitivity, frequency range, calibration accuracy, and duty cycle of GW detectors.
    \item Ensure that all appropriate observatories have responsive target-of-opportunity programs and provide a sufficient fraction of observing time to such programs.  Observatories should limit restrictions on the execution of these programs and work toward automation when feasible.
    \item Incentivize and support building tools to enable data discovery and sharing \citep[see also Chang et~al.\ white paper]{Allen19}.
    \item Develop and maintain sensitive IR instruments for photometry and spectroscopy on medium- and large-aperture telescopes on the ground and in space.
    \item Construct ELTs, which will be critical for obtaining data at late times and for lower-luminosity/distant events, in both hemispheres (see also Chornock et~al.\ white paper).
    \item Continue to support and operate sensitive $\gamma$-ray and X-ray observatories.
    \item Plan and develop new UV space observatories, with an emphasis on improving the sensitivity over current capabilities.
    \item Continue to support and improve sensitive radio observatories (e.g., ngVLA).
    \item Support the application of theory and high-performance computing for the simulations of mergers, SN explosions, and other exciting objects to connect their GW and EM signals.
    \item Obtain complementary observational (e.g., NICER) and laboratory data and perform additional theoretical modeling of the NS EoS.
    \item Perform additional laboratory measurements to improve the atomic spectral-line database of heavy elements.
    \item Enable continuous GW monitoring to ensure detection of the next Galactic SN.
\end{itemize}

\newpage

\bibliographystyle{abbrv}
\bibliography{astro_refs}

\end{document}